\documentclass[prd,aps,showpacs,preprintnumbers,11pt]{revtex4}
\usepackage{epsfig}
\usepackage{subfigure}
\begin{document}  
\newcommand{\be}{\begin{equation}}\newcommand{\ee}{\end{equation}}
\newcommand{\bea}{\begin{eqnarray}}\newcommand{\eea}{\end{eqnarray}}
\newcommand{\bc}{\begin{center}}\newcommand{\ec}{\end{center}}
\def\no{\nonumber}
\def\eq#1{Eq. (\ref{#1})}\def\eqeq#1#2{Eqs. (\ref{#1}) and  (\ref{#2})}
\def\lsim{\raise0.3ex\hbox{$\;<$\kern-0.75em\raise-1.1ex\hbox{$\sim\;$}}}
\def\gsim{\raise0.3ex\hbox{$\;>$\kern-0.75em\raise-1.1ex\hbox{$\sim\;$}}}
\def\slash#1{\ooalign{\hfil/\hfil\crcr$#1$}}
\def\eff{\mbox{\tiny{eff}}}
\def\order#1{{\mathcal{O}}(#1)}
\def\pppm{B^0\to\pi^+\pi^-}
\def\pzpz{B^0\to\pi^0\pi^0}
\def\pppz{B^0\to\pi^+\pi^0}
\preprint{}
\title{Pion as a Longitudinal Axial-Vector Meson $q\bar{q}$ Bound State}
\author{T. N. Pham }
\affiliation{
Centre de Physique Th\'{e}orique, CNRS,
Ecole Polytechnique, 91128 Palaiseau, Cedex, France}

\date{\today}
\begin{abstract}
The success of the Adler-Bell-Jackiw(ABJ)  chiral anomaly prediction for
 $\pi^{0}\to \gamma\gamma$ decay rate shows that non-anomaly terms would make
a  negligible contribution to the decay rate, in agreement
 with the Sutherland-Veltman theorem. Thus the conventional
 $q\bar{q}$ bound-state description  of the pion could not be valid since it 
would produce a   $\pi^{0}\to \gamma\gamma$ decay amplitude not suppressed
in the soft pion limit, in contradiction with the Sutherland-Veltman  theorem.
Therefore, if the pion is to be treated as a $q\bar{q}$ 
bound state, this bound state would be a longitudinal axial-vector meson.
In this paper, we consider   the pion to be a longitudinal axial-vector
 meson $q\bar{q}$ bound state with derivative coupling for
the pion  $q\bar{q}$ Bethe-Salpeter(BS) wave function. We shall 
show that this BS wave function could produce a suppressed 
$\pi^{0}\to \gamma\gamma$ decay amplitude in the soft pion limit, in agreement
with the Sutherland-Veltman theorem. This explains the almost perfect
agreement of  the anomaly
prediction with experiment and the suppression of the
virtual one-photon exchange contribution in  $\eta\to 3\pi$ decay. The
Goldstone boson equivalence theorem used  for longitudinal gauge bosons
scattering in the electroweak standard model then identifies 
the longitudinal axial-vector meson $q\bar{q}$ bound state
 with the pion.

\end{abstract}
\maketitle

\section{Introduction}
Chiral symmetry, as known from the success of the 
Goldberger-Treiman relation for the pion-nucleon coupling
constant obtained from the PCAC hypothesis, is a good symmetry
of strong interactions. The  spontaneous breakdown 
of the $SU(2)\times SU(2)$ chiral symmetry generates a massless
 Nambu-Goldstone boson which then acquires a small mass through a chiral
symmetry breaking quark mass term. PCAC and  Adler-Bell-Jackiw 
 chiral anomaly \cite{Adler,Adler1,Bell} then produce the 
$\pi^{0}\to \gamma\gamma$ decay rate in good agreement with experiment.
On the other hand, in a conventional bound-state model, a neutral
pseudoscalar $q\bar{q}$ $0^{-+}$ state, like the  $\eta_{c}$ meson,
 is  usually  massive and could decay into two photons like the two-photon
 decays of  positronium and heavy quarkonium. Being massive,
 they cannot be identified with   the neutral pseudoscalar meson of the 
ground state $SU(3)$ octet like  $\pi^{0}$ and  $\eta$ meson, 
the  Nambu-Goldstone bosons of the $SU(3)\times SU(3)$ chiral symmetry. In 
the traditional non-relativistic and relativistic bound-state calculations,
one could compute the $\pi^{0}\to \gamma\gamma$ decay rate  
using the physical pion mass and obtains some agreement with 
experiment \cite{Haye,Isgur,Barnes,Resag}, but this  
particle could not be the pion,  since the two-photon
decay amplitude for this pseudoscalar  $q\bar{q}$ state is not suppressed
in the soft pion limit according to the Sutherland-Veltman 
theorem \cite{Sutherland,Veltman}. The pion could however be in a longitudinal
 axial-vector meson $q\bar{q}$ state, if this state could produce
  a suppressed $\pi^{0}\to \gamma\gamma$ decay amplitude in the
 soft pion limit so that the agreement with experiment for the ABJ 
anomaly prediction of the $\pi^{0}$ two-photon decay rates is preserved.
 In  this paper, we shall show that, with
 the  longitudinal axial-vector meson pion BS wave function, the 
$\pi^{0}\to \gamma\gamma$ decay amplitude would be  suppressed in the soft
pion limit, in agreement with the Sutherland-Veltman theorem. This
explains the almost perfect agreement of  the anomaly
prediction with experiment for $\pi^{0}\to \gamma\gamma$ decay
and  the suppression of the virtual one-photon exchange contribution
 in  $\eta\to 3\pi$ decay which  is then given, to leading order
 in Chiral Perturbation Theory, by the non-electromagnetic isospin
 breaking current quark mass term of the QCD Lagrangian of the standard model.
\section{The Sutherland-Veltman Theorem}

 Since the basis of our analysis is the Sutherland-Veltman theorem, for
convenience, we  reproduce  this theorem here. Writing the
  $\pi^{0} \to  \gamma\gamma$ amplitude  in the original 
notation \cite{Sutherland}, we have:
\be
 g\,\epsilon_{\alpha\beta\gamma\delta}\epsilon_{1\alpha}\epsilon_{2\beta}
k_{1\gamma}k_{2\delta} = \epsilon_{1\alpha}\epsilon_{2\beta}\int\, <0|{\rm T}[j_{1\alpha}(x)\,j_{2\beta}(0)|\pi^{0}_{q}>\exp{(-i\,k_{1}\cdot x)} d^{4}x.
\label{Agg}
\ee
Using PCAC with
\be
 \partial_{\mu}A^{\mu}= f_{\pi}m_{\pi}^{2}\varphi_{\pi}, 
\label{PCAC}
\ee
one finds:
\bea
&&\frac{(q^{2}-
  m_{\pi}^{2})}{f_{\pi}m_{\pi}^{2}}\,\epsilon_{1\alpha}\epsilon_{2\beta}\int\, <0|{\rm T}[j_{1\alpha}(x)\,j_{2\beta}(0)\partial_{\mu}j_{\mu}^{5}(z)|0>\,\exp{(-i\,k_{1}\cdot x+i\,q\cdot z)} d^{4}x\,d^{4}z  \nonumber \\
&& = \frac{(q^{2}-
  m_{\pi}^{2})}{f_{\pi}m_{\pi}^{2}}\,\epsilon_{1\alpha}\epsilon_{2\beta}\,q_{\mu}\int\, <0|{\rm T}[j_{1\alpha}(x)\,j_{2\beta}(0)j_{\mu}^{5}(z)|0>\, \exp{(-i\,k_{1}\cdot x+i\,q\cdot z)} d^{4}x\,d^{4}z.
\label{SVT}
\eea 
Since gauge invariance requires that
\be
\int\, <0|{\rm
  T}[j_{1\alpha}(x)\,j_{2\beta}(0)j_{\mu}^{5}(z)|0>\,\exp{(-i\,k_{1}\cdot x+i\,q\cdot z)} d^{4}x\,d^{4}z \propto \epsilon_{\alpha\beta\nu\sigma}k_{1\nu}k_{2\sigma}q_{\mu}. 
\label{GI}
\ee
for  $q^{2}=0$ ($q$ being the pion momentum), $g\to 0$
in the soft pion limit, 
the amplitude  $\pi^{0}\to \gamma\gamma$ is
 $O(q^{2})$ and becomes suppressed as $q^{2}\to 0$.  This theorem is
 now evaded by the ABJ the chiral anomaly in the quark triangle graph
 which gives us the well-known chiral anomaly prediction 
for $\pi^{0} \to \gamma\gamma$ decay. To see how this comes about, we 
reproduce here \cite{Pham1,Pham2}the derivation of the $\pi^{0} \to
\gamma\gamma$ amplitude using the modified  PCAC equation: 
\be
\partial_{\mu}A^{\mu}= f_{\pi}m_{\pi}^{2}\phi + S\frac{e^{2}}{16\pi^{2}}
\epsilon_{\alpha\beta\gamma\delta}F^{\alpha\beta}F^{\gamma\delta}
\label{dA}
\ee
with $S=1/2$ in the standard model.
Taking the matrix element of both sides of Eq. (\ref{dA}) and separating
the $\pi^{0}$ pole term, we find, in the notation of \cite{Pham1,Pham2}:
\be
N^{\mu\nu} = \frac{1}{f_{\pi}}\left( p_{\tau}\tilde{R}^{\mu\nu\tau}(q,k)
- S\frac{e^{2}}{2\pi^{2}}Y^{\mu\nu}\right)
\label{Npi}
\ee
with $<\pi^{0}(p)|T|\gamma^{*}(q)\gamma(k)>=
\epsilon_{\mu}(q)\epsilon_{\nu}(k)N^{\mu\nu}(q,k) $ and 
$N^{\mu\nu}(q,k) $ given by:
\be
N^{\mu\nu}(q,k)= e^{2}F(q,k)Y^{\mu\nu}, \quad Y^{\mu\nu}= 
\epsilon^{\mu\nu\alpha\beta}q_{\alpha}k_{\beta}.
\label{amp}
\ee
where $F(q,k)$ is the transition form factor and $p,k,q$ are
respectively, the pion and the two photons momenta. In Eq. (\ref{Npi})
 $\tilde{R}^{\mu\nu\tau}(q,k) $ is 
the triangle graph  (the direct coupling between the three currents)
or the continuum contribution to the axial vector current matrix element
$<0|A_{\mu}|\gamma^{*}\gamma>$ defined as $R^{\mu\nu\tau}(q,k) $:
\be
R^{\mu\nu\tau}(q,k) = \tilde{R}^{\mu\nu\tau}(q,k)  - f_{\pi}
\frac{p^{\tau}N^{\mu\nu}(q,k)}{p^{2}-m_{\pi}^{2}}
\label{Rmnt}
\ee
 Gauge invariance and Bose symmetry tells us
that when both photons are real as in the  $\pi^{0}\to \gamma\gamma$
 decay ($q^{2}= 0, k^{2}= 0$),  the divergence 
$p_{\tau}R^{\mu\nu\tau}(q,k) $ is  $O(p^{2})$ and becomes negligible.
 One can then apply Eq. (\ref{Npi}) to the $\pi^{0}\to \gamma\gamma$ 
decay amlitude and finds that it is given by the
 anomaly \cite{Adler,Adler1,Bell}. From the expressions in Ref. \cite{Adler,Rosenberg}, the triangle graph contribution to the direct term is then
\be
 p_{\tau}\tilde{R}^{\mu\nu\tau}(q,k) = e^{2}S\biggl(2mP(q,k) + \frac{1}{2\pi^{2}}\biggr)Y^{\mu\nu}
\label{Rt}
\ee 
with 
\be
 P(q,k) =
 \frac{m}{2\pi^{2}}\int_{0}^{1}dx\int_{0}^{1-x}\frac{dy}{[k^{2}y(1-y) +
   q^{2}x(1-x) -(q^{2} + k^{2}- p^{2})xy -m^{2}]}
\label{Pm}
\ee 
When both photons are real ($q^{2}=0,k^{2}=0$), from  Eq. (\ref{Pm}), 
we get:
\be
2mP(q,k) =  -\frac{1}{2\pi^{2}} + O(p^{2})
\label{Pm0}
\ee 
which implies that the r.h.s of Eq. (\ref{Rt}) is $O(p^{2})$
in agreement with our previous remark that
$p_{\tau}\tilde{R}^{\mu\nu\tau}(q,k) = O(p^{2})$ which is precisely the 
Sutherland result for the axial vector current matrix element
 in Eq. (\ref{GI}). However
the PCAC equation has been modified by the triangle graph anomaly which
gives rise to  the  second term in  Eq. (\ref{Npi}) from which
 we get the anomaly prediction for  $\pi^{0}\to \gamma\gamma$,
\be
 A(\pi^{0} \to \gamma\gamma)=
 \epsilon^{\mu\nu\alpha\beta}\,k_{\alpha}k^{\prime}_{\beta}\,
\biggl(- \frac{e^{2}}{4\pi^{2}f_{\pi}}\biggr).
\label{p2gg}
\ee

This result is exact to all order in  $\alpha$ and
the strong coupling constant $\alpha_{s}$ and is in almost 
perfect agreement with experiment, confirming
the validity of the soft-pion theorem in the presence of the ABJ
anomaly. Other non-anomaly contribution is $O(p^{2})$ and is suppressed
in agreement with the Sutherland-Veltman theorem as mentioned above.

It is then quite surprising to find in the literature paper claiming 
to reproduce exactly the anomaly prediction using the impulse 
 approximation  \cite{Roberts} to calculate the width for 
$\pi^{0} \to \gamma\gamma$ decay. It is evident that with an  approximation, 
one cannot expect to reproduce exactly the anomaly
 prediction of Adler \cite{Adler},  since various corrections have
 to be included, thus invalidating the result of Ref. \cite{Roberts}
 which appears as pure numerology, in our opinion.
Since  the anomaly prediction is an exact  model-independent soft-pion
result,  like the low-energy theorem for Compton scattering on any
 target , the Thomson formula  $-(e^{2}/m)(e_{1}\cdot e_{2})$, it 
cannot be obtained by any  model calculation. Any model 
calculation  of $\pi^{0} \to \gamma\gamma$ decay amplitude in QCD
using BS wave function  will get  $\alpha_{s}$ corrections and
errors from neglecting other contributions,  and in the calculation of 
\cite{Roberts}, there is no physical principle to protect
the result from these $\alpha_{s}$ corrections and other
contributions  neglected in the impulse approximation. Moreover the use of a 
momentum-independent BS wave function for the pion, the $\gamma_{5}$
term,  will produce  terms not suppressed in the soft-pion limit, for
 both the non-anomaly term in $\pi^{0} \to \gamma\gamma$ and the
contribution from virtual one-photon exchange electromagnetic
interactions in  $\eta\to 3\pi$ decay amplitude as mentioned
below. Therefore the calculation of \cite{Roberts} is incorrect.  The
 fallacy in  that work is the use of an approximation for an exact result. 

At the same time with the Sutherland-Veltman theorem for 
$\pi^{0} \to \gamma\gamma$, Sutherland \cite{Sutherland,Sutherland1}  
and Adler \cite{Adler2}  also  show that the $\eta\to 3\pi$ decay
 is suppressed in the soft-pion limit. This is because the vanishing
 of the E. T. C. in the soft-pion expression
implies that the $\eta\to 3\pi$ decay with $q\to 0$ (q being the $\pi^{0}$
momentum) is suppressed. The reason is that, as  shown  by 
Cantor \cite{Cantor}, for the virtual one-photon exchange
electromagnetic  interactions which transform  as 
$(\underline{1},\bar{\underline{8}}) + (\bar{\underline{8}},\underline{1})$
representation of $SU(3)\times SU(3)$, only derivative coupling
is  allowed in the  $\eta\to 3\pi$ decay amplitude. In fact a  suppressed
 $\eta-\pi^{0}$ mixing could  be obtained easily from the virtual one-photon
 exchange electromagnetic interactions if one uses the
 momentum-dependent BS wave function, in agreement with the
Sutherland theorem \cite{Sutherland1,Adler2}. One thus needs a new
non-electromagnetic isospin violating tadpole term to obtain the 
unsuppressed $\eta\to 3\pi$ decay rate. For this the tadpole term,  
Cantor \cite{Cantor} adds a non-electromagnetic isospin breaking term $u_{3}$ to the
Gell-Mann-Oakes-Renner  $(\bar{\underline{3}},\underline{3})$
 $SU(3)\times SU(3)$ breaking term \cite{Gell-Mann}:
\be
\mathcal{ L}_{1} = a_{0}u_{0} + a_{8}u_{8} + a_{3}u_{3}
\ee
The idea that this $u_{3}$ term is essential for $\eta\to 3\pi$ decay
is due to Cabibbo and Wilson as quoted by Cantor \cite{Cantor}. The
soft-pion theorem for the matrix element $<\eta|u_{3}|3\pi>$ then 
produces  the non-derivative term for $\eta\to 3\pi$ decay which is not
suppressed in the soft-pion limit. This is the origin of the
 non-electromagnetic isospin breaking current quark mass term of the
QCD Lagrangian which gives, at leading order in Chiral Perturbation 
Theory, the $\eta\to 3\pi$ decay amplitude.

\section{Pion as a Longitudinal Axial-Vector Meson}

We have seen that without the ABJ anomaly, the $\pi^{0} \to \gamma\gamma$
decay would be suppressed. In any model calculation, for
example, in  non-relativistic or relativistic calculation, without 
PCAC and chiral symmetry, the Sutherland-Veltman theorem does not apply
and the two-photon decay is not suppressed in the soft pion limit as
found in existing bound-state calculations of quarkonium two-photon
decays \cite{Haye,Isgur,Barnes,Resag}. The suppression of the
virtual photon exchange electromagnetic interactions in  $\eta\to 3\pi$
decay is another chiral symmetry constraint to be imposed on the pion BS
wave function according to Sutherland theorem for $\eta\to 3\pi$ decay.

 It follows that the pion 
could not  be described by the usual  
momentum-independent $q\bar{q}$ bound-state wave function. Since many 
properties of hadrons, and  in particular, the light mesons 
and quarkonium systems, are well described by the $q\bar{q}$ bound-state
picture, the problem  is how to reconcile this bound-state picture 
with the Nambu-Goldstone boson character of the pion. The solution of
the problem could be found easily by looking at the solution of the
relativistic bound-state Bethe-Salpeter(BS) equation \cite{Bethe}
for a $q\bar{q}$ system. For a pseudoscalar meson, there are two possible
solutions. The  solution with the momentum-independent
wave function of the form $P\gamma_{5} $ and the longitudinal
axial-vector momentum-dependent ${\not p}\gamma_{5}A $ solutions. This
longitudinal solution  has been considered by Kummer \cite{Kummer}.  
The longitudinal axial-vector meson wave function for pion is also
 used  by Chernyak and Zhitnitsky for process involving pion
 at high energy \cite{Chernyak}. 

 As mentioned above, the $P\gamma_{5}$  solution would be 
 in contradiction with the  Sutherland-Veltman theorem and therefore
 could not  be the correct  pion $q\bar{q}$ bound-state wave function. 
The ${\not p}\gamma_{5}A $ solutions  would be acceptable.  In fact, 
if the pion is a longitudinal axial-vector meson $q\bar{q}$ bound state, the 
 $\pi^{0} \to \gamma\gamma$ amplitude computed with this
wave function, as shown below, would be similar to the free quark triangle
graph contribution to the two-photon matrix element of the axial-vector
current divergence $<0|\partial_{\mu}A_{\mu}(0)|\gamma\gamma> $ and
 therefore vanishes in the massless quark limit and
thus does not contribute to the $\pi^{0} \to \gamma\gamma$ decay.
In the following we present a computation of the 
$\pi^{0} \to \gamma\gamma$ amplitude using the longitudinal axial-vector
meson as the pion BS wave function. Consider now  the BS wave function
 of \cite{Jain}:
\be
\psi(p,q)= \gamma_{5}\!\,\psi_{0} + \gamma_{5}\!\not{p}\psi_{1} + 
\gamma_{5}\!\not{q}p\cdot q\psi_{2}+\gamma_{5}\!\,[\not{q},\not{p}]\psi_{3}. 
\label{BS}
\ee
where $p$ and  $q$ is the pion and relative momentum of the $q\bar{q}$
system, with the quark and anti-quark momwentum
 $q_{1}= q + p/2$, $q_{2}= q -p/2$ and  $\psi_{i}, i=0,...,3$
are the scalar functions of $p$ and  $q$.  The first term $\psi_{0}$
in  Eq. (\ref{BS}) is the momentum-independent wave function, as mentioned 
above,  produce a  $\pi^{0} \to \gamma\gamma$ decay 
in the soft pion limit and is dropped here. The third term $\psi_{2}$ 
which is $O(p\cdot q)$ could give a contribution $O(p)$ in the soft pion
 limit and need not to be considered here. The
last term $\psi_{3}$, does not make a contribution  to  $\pi^{0} \to
\gamma\gamma$ decay by the triangle graph. This leaves us 
with the $\psi_{1}$ term as the 
longitudinal contribution to the $\pi^{0} \to \gamma\gamma$
decay. The BS equation \cite{Jain}  for $ \psi(p,q)$ with the  gluon
 propagator $G_{\mu\nu}(k-q)$  reads:
\be
(\not{q}+\not{p}/2)\psi(p,q)(\not{q}-\not{p}/2) = -i\int \frac{d^{4}q^{\prime}}{(2\pi)^{4}} \gamma_{\mu}\psi(p,q^{\prime})\gamma_{\nu}G_{\mu\nu}(q^{\prime}-q).
\label{BSE}
\ee
Since, by definition, the BS vertex function $\Gamma(p,q)$ is the BS wave
function with the free quark propagator removed \cite{Pham,Resag1},
 Eq. (\ref{BSE}) can be used to express $\Gamma(p,q)$ in terms of the BS
wave function  $\psi(p,q)$. We have:
\be
\Gamma(p,q) =  -i\int\frac{d^{4}q^{\prime}}{(2\pi)^{4}} \gamma_{\mu}\psi(p,q^{\prime})\gamma_{\nu}G_{\mu\nu}(q^{\prime}-q).
\label{BSV}
\ee
\begin{figure}[h]
\centering
\includegraphics[height=4.0cm,angle=0]{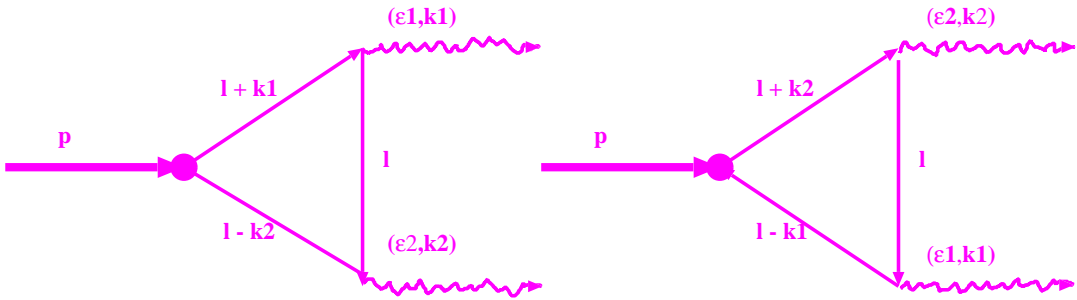}
\caption{  Quark loop triangle graphs with BS longitudinal axial-vector
meson wave  function\\  for $\pi^{0}\to \gamma\gamma$ decay}.
\label{pi2gg}
\end{figure}

In the following, as our purpose is to obtain the soft pion limit for
$\pi^{0}\to \gamma\gamma$ decay, we consider only  the longitudinal
 solution for the BS wave function $\gamma_{5}\!\not{p}\,\psi_{1} $ given
 in Eq. (\ref{BS}), and  for simplicity, we use the gluon propagator
in the Feynman gauge with
  $G{\mu\nu}(q^{\prime}-q)=-g_{\mu\nu}/(q^{\prime}-q)^{2} $.
The $\pi^{0}\to \gamma\gamma$ decay amplitude is given by the
 quark loop triangle graph similar to  the ABJ chiral anomaly
 triangle graph, except that the point-like axial-vector current 
vertex $\gamma_{\mu}\gamma_{5}$ is replaced by the BS longitudinal
 axial-vector meson wave function 
$\psi(p,q^{\prime})=\gamma_{5}\!\not{p}\,\psi_{1}(p,q^{\prime})$, and the  
factor $1/(q^{\prime}-q)^{2}$ from the gluon propagator which makes the
integration over $q$ convergent and could be carried out by the
usual change of variable, assuming the integral over $q^{\prime}$
convergent. Similar to the calculation of Ref. \cite{Adler}, the 
$\pi^{0}\to \gamma\gamma$ decay amplitude with the BS vertex 
function $\Gamma(p,q)$ shown in Fig. \ref{pi2gg}, after a change of variable
 $l= q + p/2$, with $l$ one of the quark momentum in the triangle 
loop and  $\Gamma(p,q)=\Gamma(p,l)$ and putting $m=0$, is given by:
\be
M = -ie^{2}\int\frac{d^{4}l}{(2\pi)^{4}}\rm{Tr}\biggl(\not{\epsilon_{2}}\frac{1}{(\not{l}-  \not{k_{2}})}\not{p}\gamma_{5}\frac{1}{(\not{l}+
  \not{k_{1}})}\not{\epsilon_{1}}\frac{1}{\not{l}}\biggr)\,J(p,l)  + (\epsilon_{1},k_{1}
\to \epsilon_{2},k_{2}\  \rm{terms} )
\label{Api2gg}
\ee
with the scalar part of the BS vertex function $\Gamma(p,l)$ given by:
\be
J(p,l) =  -2\int\frac{d^{4}l^{\prime}}{(2\pi)^{4}}\frac{\psi_{1}(p,l^{\prime})}{(l^{\prime}-l)^{2}}.
\label{J}
\ee
Using the identity \cite{Adler},
\be
\frac{1}{(\not{l}-  \not{k_{2}})}\not{p}\gamma_{5}\frac{1}{(\not{l}+
  \not{k_{1}})} = \frac{1}{(\not{l}-  \not{k_{2}})}\gamma_{5} + \gamma_{5}
\frac{1}{(\not{l}+  \not{k_{1}})} ,\quad  p=k_{1} + k_{2}
\label{Id}
\ee
The Dirac $\gamma$ term (the Trace term), is then split into two
 contributions.  The contributions from the
1st and 2nd diagrams in Fig. \ref{pi2gg} are respectively then:
\bea
&& T_{1}= \frac{\not{\epsilon_{2}}(\not{l}-  \not{k_{2}})\not{\epsilon_{1}}\!
\not{l}\gamma_{5}}{(l-k_{2})^{2}l^{2}} -\frac{\not{\epsilon_{2}}\!(\not{l}+ \not{k_{1}})\not{\epsilon_{1}}\!
\not{l}\gamma_{5}}{(l+k_{1})^{2}l^{2}} \label{T1}\\
&& T_{2}= \frac{\not{\epsilon_{1}}(\not{l}-  \not{k_{1}})\not{\epsilon_{2}}\!
\not{l}\gamma_{5}}{(l-k_{1})^{2}l^{2}} -\frac{\not{\epsilon_{1}}(\not{l}+ \not{k_{2}})\not{\epsilon_{2}}\!\not{l}\gamma_{5}}{(l+k_{2})^{2}l^{2}} 
\label{T2}
\eea
We see that, provided that the integral over $l$ converges,
 the $k_{2}$-terms in Eq. (\ref{T1}) and Eq. (\ref{T2})
would cancel after integration over $l$ by a change of variable
$l -k_{2}\to l$ and $l \to l + k_{2}$ in the $k_{2}$-terms of Eq. (\ref{T1})
and similarly for the $k_{1}$-terms
with a change of variable $l -k_{1}\to l$ and $l \to l + k_{1}$ 
in  Eq. (\ref{T2}). This is not the case 
with  point-like axial-vector current in the triangle graph since
the shift of the integration variable $l -k_{2}\to l$ in Eq. (\ref{T1}),
or $l -k_{1}\to l$ in Eq. (\ref{T2}), would induce an anomaly 
term \cite{Peskin}. This is the well-known anomaly terms for the
divergence of the axial-vector current \cite{Adler}. In our bound-state
calculation, the point-like axial-vector current is replaced 
by the longitudinal axial-vector meson  BS vertex function and 
the $1/l^{2}$ behavior of the gluon propagator at large $l^{2}$
would make the integrals over $l$ convergent for  $k_{1}$ and  $k_{2}$
terms in the two diagrams. Taking the trace, the total contribution to 
$\pi^{0}\to \gamma\gamma$ decay amplitude is then given by:
\be
M = -i\,e^{2}\int\frac{d^{4}l}{(2\pi)^{4}}\biggl( -\frac{4i\epsilon(\epsilon_{1},\epsilon_{2},k_{1},l)4l\cdot k_{1}}{(l^{2}(l-k_{1})^{2}(l+k_{1})^{2})} + \frac{4i\epsilon(\epsilon_{1},\epsilon_{2},k_{2},l)4l\cdot k_{2}}
{(l^{2}(l-k_{2})^{2}(l+k_{2})^{2})}\biggr)J(p,l).
\label{Mpi2gg}
\ee
where $\epsilon(\epsilon_{1},\epsilon_{2},k_{1},l)$ and 
$\epsilon(\epsilon_{1},\epsilon_{2},k_{2},l)$ denote the contraction of
 $\epsilon_{1},\epsilon_{2},k_{1},l$ and $\epsilon_{1},\epsilon_{2},k_{2},l$
with the anti-symmetric tensor $\epsilon$. Assuming that the integral over 
$l^{\prime}$ in $J(p,l)$ is finite, the integration over $l$
in the above expression will produce terms proportional to
$\epsilon(\epsilon_{1},\epsilon_{2},k_{1},k_{1})$,
$\epsilon(\epsilon_{1},\epsilon_{2},k_{1},l^{\prime})\,l^{\prime}\cdot k_{1}$
for $k_{1}$-term and $\epsilon(\epsilon_{1},\epsilon_{2},k_{2},k_{2})$, 
$\epsilon(\epsilon_{1},\epsilon_{2},k_{2},l^{\prime})\,l^{\prime}\cdot k_{2}$
for $k_{2}$-term in Eq. (\ref{Mpi2gg}). Since
$\epsilon(\epsilon_{1},\epsilon_{2},k_{1},k_{1})=0$ , $\epsilon(\epsilon_{1},\epsilon_{2},k_{2},k_{2})=0$, only the $l^{\prime}$
term survives  after integration over $l$.
After integration over $l^{\prime}$, only terms proportional to
$p\cdot k_{1}$  and $p\cdot k_{2}$ survive, but these are  $O(p^{2})$
and are  suppressed in the soft pion limit, in agreement with the
 Sutherland-Veltman theorem. Provided that the integrals over $l$ and 
$l^{\prime}$ are finite, this result does not depend on the detailed form of
the BS wave function and the use of the one-gluon exchange kernel in $J(l,p)$.
 The $\pi^{0}\to \gamma\gamma$ decay is 
then given by the ABJ anomaly which agrees well with experiment. This
implies the absence of the  $P\gamma_{5}$ term in the pion
BS wave function and the pion thus behaves as a longitudinal
axial-vector meson. This  momentum-dependent  pion  BS wave function
 will produce  a 
suppression of the $\eta\to 3\pi$ decay in agreement with Sutherland
theorem. 

  In the electroweak standard
model, the unphysical Goldstone boson becomes the longitudinal gauge boson,
in our case, the longitudinal axial-vector meson   $q\bar{q}$ bound
state appears as a Goldstone boson, the pion,  according to the 
 Goldstone boson equivalence theorem for   $W_{L}W_{L}$
 scattering in the standard model \cite{Donoghue}. Since only the
 kinetic term is generated with this BS wav function,  pion remains
 massless.  The pion mass term
has to be generated by  chiral symmetry breaking term as the $\sigma$-term 
in the $\sigma$ model \cite{Gell-Mann2} which is now identified as the
quark mass term in the QCD Lagrangian of the standard model. 

\section{Conclusion}
   In conclusion, we have derived the Sutherland-Veltman theorem for 
the $\pi^{0}\to \gamma\gamma$ decay considering  the pion as a longitudinal
 axial-vector meson $q\bar{q}$ bound state. With this longitudinal
 axial vector meson BS wave we have been able to obtain the suppressions
 of the non-anomaly term in $\pi^{0}\to \gamma\gamma$  decay and the 
 virtual one-photon exchange electromagnetic interaction term in
$\eta \to 3\pi$ decay.  The Goldstone boson equivalence theorem 
used  for $W_{L}W_{L}$ scattering in the electroweak standard model 
then identifies  the longitudinal axial-vector meson $q\bar{q}$
 bound state with the pion.  This allows us
 to say that the pion could be a $q\bar{q}$ bound state at the same
 time a Nambu-Goldstone boson  of chiral symmetry with the two-photon
 decay given by PCAC and the ABJ chiral anomaly.  The momentum-dependent
 BS wave function could then be used to obtain the derivative couplings
 with hadrons, in agreement with chiral symmetry.


\begin{thebibliography}{99}

\bibitem{Adler} S. L. Adler,  Phys. Rev. {\bf  177}, 2426 (1969).

\bibitem{Adler1} S. L. Adler, {\em Lectures on Elementary particles and 
quantum field theory}, Brandeis University Summer Institute in Theoretical
Physics,  eds. S. Deser, M. Grisaru and H. Pendleton
 (MIT Press, Cambridge, MA).

\bibitem{Bell} J. S. Bell and R. Jackiw, Nuovo Cimento {\bf 60A}, 37 (1969).

\bibitem{Haye} C. Haye and N. Isgur,  Phys.\ Rev.\ D {\bf 25}, 1944 (1982).

\bibitem{Isgur} S. Godfrey and N. Isgur,  Phys.\ Rev.\ D {\bf 32},
 189 (1985).

\bibitem{Barnes} E. S. Ackleh and T. Barnes,  Phys.\ Rev.\ D {\bf 45},
 232 (1992).

\bibitem{Resag}  C. R. M\"unz, J. Resag,  B. C. Mestsch, and H. R. Petry
 Nucl. Phys. \ A {\bf 578}, 418 (1994).

\bibitem{Sutherland} D. G. Sutherland,  Nucl. Phys. {\bf B 2}, 433 (1967).

\bibitem{Veltman} M. Veltman,  Proc. R. Soc. {\bf 301A}, 107 (1967).

\bibitem{Rosenberg} L. Rosenberg,  Phys.\ Rev. {\bf 129} (1963) 2786.


\bibitem{Roberts} C. D. Roberts,  Nucl. Phys. A {\bf 605}, 14 (1996).

\bibitem{Sutherland1} D. G. Sutherland,  Phys.\ Lett.\ B {\bf 23}
384 (1966).

\bibitem{Adler2} S. Adler, Phys.\ Rev. Lett. {\bf 18}, 519  (1967).


\bibitem{Pham1} T. N. Pham  and X. Y. Pham, Phys.\ Lett.\ B {\bf 247}
438 (1990). 

\bibitem{Pham2} T. N. Pham  and X. Y. Pham, Int. J. Mod. Phys. {\bf A26}, 4125 (2011).

\bibitem{Cantor} A. J. Cantor,  Phys.\ Rev. D {\bf 3}, 3195, 3205 (1971).

\bibitem{Gell-Mann} M. Gell-Mann, R. J. Oakes and B. Renner, 
 Phys.\  Rev.  {\bf 175}, 2195 (1968).

\bibitem{Bethe} E. E. Salpeter and H. A. Bethe,  Phys.\ Rev. {\bf 84},
 1232  (1951).

\bibitem{Kummer} W. Kummer, Nuovo Cim. {\bf 31}, 219 (1964). 

\bibitem{Chernyak}  V. I. Chernyak  and A. R. Zhitnitsky, 
  Nucl. Phys.\ B {\bf 201}, 492 (1982).


\bibitem{Jain} P. Jain and H. J. Munczek,  Phys.\ Rev.\ D {\bf 44},
 1873 (1991).

\bibitem{Pham} T. N. Pham,  Phys.\ Rev.\ D  {\bf 19}, 707 (1979).

\bibitem{Resag1}  J. Resag, C. R. M\"unz, B. C. Mestsch, and H. R. Petry
 Nucl. Phys. \ A {\bf 578}, 397 (1994).

\bibitem{Peskin} M. E. Peskin and D. V. Schroeder, {\em An Introduction
 to {\bf QUANTUM FIELD THEORY}},\\ Addison-Wesley
 Publishing Company, New York (1995).

\bibitem{Donoghue} See, e.g. J. F. Donoghue, E. Golowich, and B. R. Holstein, 
{\em Dynamics of the Standard Model},\\ Cambridge University Press, Cambridge
(1992).

\bibitem{Gell-Mann2} M. Gell-Mann and M. Levy, Nuovo Cim. {\bf 16}, 705 (1960). 





\end{thebibliography}
\end{document}